\documentclass[aps,pra,notitlepage,superscriptaddress,floatfix,10pt,nofootinbib,twocolumn]{revtex4-2}
\usepackage[utf8]{inputenc}
\usepackage[english]{babel}
\usepackage[T1]{fontenc}
\usepackage{amsmath}

\usepackage{tikz}
\usepackage{lipsum}
\usepackage{graphicx}%
\usepackage{amsmath}
\usepackage{rotating}
\usepackage{extarrows}
\usepackage{color,calc,graphicx}
\usepackage[colorlinks]{hyperref}
\hypersetup{
	colorlinks = true,
	urlcolor = {blue},
	citecolor = {magenta},
	linkcolor= {blue}
}
\usepackage{dcolumn}
\usepackage{bm}
\usepackage[all]{xy}
\usepackage{indentfirst}
\usepackage{tabularx}
\usepackage{amsmath}
\usepackage{bbm}
\usepackage{color}
\usepackage{multirow}
\usepackage{mathrsfs}
\usepackage{euscript}
\usepackage{amssymb}

\newcommand{\ket}[1]{|#1\rangle}

\newcommand{\oI}{\overline{I}}

\newcommand{\sA}{\textsf{A}}
\newcommand{\sB}{\textsf{B}}

\newcommand{\bR}{\textbf{R}}
\newcommand{\ba}{\textbf{a}}
\newcommand{\bb}{\textbf{b}}
\newcommand{\bc}{\textbf{c}}
\newcommand{\bx}{\textbf{x}}

\newcommand{\br}{\textbf{r}}

\newcommand{\btf}{\textbf{f}}
\newcommand{\bu}{\textbf{u}}

\newcommand{\cH}{\mathcal{H}}

\newcommand{\cD}{\mathcal{D}}
\newcommand{\cM}{\mathcal{M}}

\DeclareMathOperator\tr{Tr}
\newcommand\mean[1]{\left\langle #1 \right\rangle}

\def\eqabove#1  {\stackrel{\mathclap{\footnotesize\mbox{#1} }} {=}  }%
\def\leqabove#1 {\stackrel{\mathclap{\normalfont\mbox{#1} }} {\leq}  }%
\begin{document}
\title{Experimental Characterization of Quantumness Using the Uncertainty Principle, Coherence, and Nonlocality}
\author{Yan-Han Yang}
\author{Xin-Zhu Liu}
\thanks{These authors contribute equally.}

\author{Xing-Zhou Zheng}
\author{Jun-Li Jiang}

\author{Xue Yang}
\affiliation{School of Information Science and Technology, Southwest Jiaotong University, Chengdu 610031, China}

\author{Shao-Ming Fei}
\email{feishm@cnu.edu.cn}
\affiliation{School of Mathematical Sciences, Capital Normal University, Beijing 100048, China}

\author{Zhihao Ma}
\email{mazhihao@sjtu.edu.cn}
\affiliation{School of Mathematical Sciences,  MOE-LSC, Shanghai Jiao Tong University, Shanghai 200240, China}
\affiliation{Mathematical Science $\&$ Digital Security Joint Library of Shanghai Seres Information Technology Co.,Ltd and School of Mathematical Science, SJTU, Shanghai 200040, China}

\author{Zizhu Wang}
\affiliation{Institute of Fundamental and Frontier Sciences, University of Electronic Science and Technology of China, 610054 Chengdu, China}

\affiliation{Key Laboratory of Quantum Physics and Photonic Quantum Information, Ministry of Education, University of Electronic Science and Technology of China, 611731, Chengdu, China}

\author{Ming-Xing Luo}
\email{mxluo@swjtu.edu.cn}
\affiliation{School of Information Science and Technology, Southwest Jiaotong University, Chengdu 610031, China}
\affiliation{Hefei National Laboratory, Hefei 230088, China}

\vspace{1cm}

\begin{abstract}
Heisenberg's uncertainty principle, coherence and Bell nonlocality have been individually examined through many experiments. In this Letter, we systematically characterize all of this quantumness in a unified manner. We first construct universal uncertainty relations to reveal intrinsic features of incompatible measurements, which include all the state-independent uncertainties as special cases.  We further extend to witness both quantum coherence and Bell nonlocality. We finally perform experiments with unified two-photon states, and validate the uncertainty principle, coherence and Bell nonlocality within the experimental error. Our methods for witnessing quantumness are valuable in characterizing quantum correlations in quantum information processing.

\end{abstract}
\maketitle

The Heisenberg Uncertainty Principle \cite{Heisenberg1927} establishes the first intrinsic uncertainty between two incompatible observables at quantum level. It imposes fundamental limitations on the precision of simultaneous measurements for pairs of incompatible quantum observables, such as position and momentum, or angle and orbital angular momentum. Initially formalized through variance-based inequalities, this principle was later generalized to bounded operators using their (anti)commutators  \cite{Kennard1927,Robertson1929,Schrodinger1930}.  These foundational insights have driven breakthroughs in various quantum technologies \cite{busch2007,Busch2013,Buscemi2014,Lu2021}.

Traditional variation-based uncertainty relations often yield trivial bounds for specific observables. To address this, there are three strategies. One is using sum-uncertainty inequalities \cite{Maccone2014,Guise2018}. The second involves leveraging constraints on measurement probability distributions, such as exploring the lower bounds of its Shannon entropy \cite{Deutsch1983,Maassen1988,Ghirardi,Vicente2008,Chen2018,UPReview}. The third method exploits Schur-concave uncertainty measures \cite{Friedland2013,Yuan2023}. These probability-based uncertainty relations extend state-dependent relations to state-independent scenarios, underpinning applications for quantum key distribution \cite{Grosshans2003}, quantum random number generation \cite{Coles2014}, entanglement detection \cite{Oppenheim2010,pre2011}, and quantum metrology \cite{Halpern2019,Xiao2023}. However, the interplay between uncertainty relations and other quantum features such as coherence and entanglement remains unclear \cite{Frank2012,Schwonnek2017}. Exhibiting their links could unveil deeper principles of quantum theory. 

\begin{figure}[!ht]
    \centering
\includegraphics[scale=0.7]{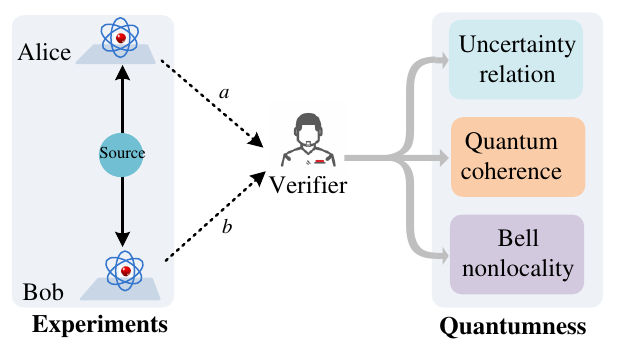}
\caption{Unified quantum game for verifying quantumness. One source prepares two copies of one particle and distributes one copy to quantum party Alice or Bob. Both players send local measurement outcomes $(a,b)$ to Verifier, who verifies specific quantum features in quantum experiment of Alice and Bob according to their measurement statistics.}
    \label{Figure1}
\end{figure}

This paper presents a unified approach to systematically characterize quantum features including uncertainty relations, coherence, and Bell nonlocality in finite-dimensional quantum systems. The main idea relies on a unified quantum game, as shown in Figure~\ref{Figure1}. We first generalize entropic uncertainty relations through a vector-value relation. We then extend this approach to exhibit both quantum coherence \cite{Baumgratz2014} and Bell nonlocality \cite{Bell1964}, which are different from previous results \cite{Oppenheim2010,Guhne2004,Zhao2018,Giovannetti2004}. We finally validate all of theoretical results through a two-photon experiment.

\textbf{Verifying state-independent uncertainty relations}. To show our main idea, we introduce state-independent uncertainty relations within a quantum game framework, shown in Fig.~\ref{Figure1}. Specifically, to verify the incompatibility of two given measurements $A$ and $B$, one source prepares two copies of a state $\rho$ drawn from a closed set $\cD \subseteq \mathbb{C}^{d}$, and sends one copy each to players, Alice or Bob. Alice performs the measurement $A=\{M_{a}\}$ on her shares and obtains outcome $a$ with probability $p_{a} \equiv \tr(M_{a}\rho)$ ($a\in[n]=\{1, \ldots, n\}$). Bob measures $B=\{N_{b}\}$ on his states, yielding outcome $b$ with probability $q_{b}\equiv\tr(N_{b}\rho)$ ($b\in[m]=\{1, \ldots, m\}$). Both outcomes $a$ and $b$ are sent to Verifier. After many rounds, Verifier can evaluate two probability distributions, $\{p_a\}$ and $\{q_b\}$, and then compute joint uncertainty (information) of each pair $(a,b)$ as $f(p_{a},q_{b})$, where $f$ is a specific analytical and bounded function serving as an informational quantity \cite{Masanes2013}. For example, inspired by Shannon entropy \cite{Cover},  one may take $f(p_a,q_b)=p_a I(p_a)+q_bI(q_b)$ with $I(p_x)=-\log p_x$, representing the uncertainty in outcomes prior to measurement. Such functions generally satisfy nonnegative, additivity, and concavity, making them well suited to quantifying the intrinsic uncertainty of incompatible measurements \cite{UPReview}. 

To witness measurement incompatibility, one interesting method is to explore state-independent uncertainty relationships. In our case, it is to bound uncertainty values  $\{f(p_{a},q_{b})\}$of a given function $f$ for all possible states, for example using entropic uncertainty relationships \cite{UPReview}.  However, these are only effective for specific choices of uncertainty functions. Instead, we propose a general framework based on vector inequalities that applies to all uncertainty functions.  Especially,  for each pair of outcomes $(a,b)$,  the quantity $f(p_{a},q_{b})$ is bounded from above and below by $\max_{\rho\in \cD}f(p_{a},q_{b})$ and $\min_{\rho\in \cD}f(p_{a},q_{b})$, respectively. To further characterize all outcomes $a\in [n]$ and $b\in[m]$, we introduce  two quantities as  
\begin{eqnarray}
&&R_{k}(A,B)
= \max_{I_{k}}\max_{\rho\in \cD}\sum_{(a',b')\in I_{k}} f(p_{a'},q_{b'}),
\\
&&r_{k}(A,B)
= \min_{I_{k}}\min_{\rho\in \cD}\sum_{(a',b')\in I_{k}}f(p_{a'},q_{b'}),
\label{uuu0}
\end{eqnarray}
where $I_{k} \subset [n] \times [m]$ is a subset of $k$ distinct outcome pairs. All uncertainty values then satisfy the following vector-value majorization inequality:
\begin{eqnarray}
\br_{\textrm{ms}}\prec\btf_{ab} \prec
\bR_{\textrm{ms}},
\label{SourceUP}
\end{eqnarray}
where all vectors are defined by 
$\bR_{ms}\equiv [R_{1}(A,B)$, $R_{2}(A,B)-R_{1}(A,B)$, $\ldots$, $R_{mn}(A,B)-R_{mn-1}(A,B)]$, $\br_{ms}\equiv [r_{1}(A,B)$, $r_{2}(A,B)-r_{1}(A,B)$, $\cdots$,  $r_{mn}(A,B)-r_{mn-1}(A,B)]$, and $\btf_{ab} \equiv [f (p_1,q_1)$, $\cdots$, $f(p_n,q_m)]$. The partial order of two vectors, i.e., $[x_1,\cdots, x_n]\prec [y_1,\cdots, y_n]$, means they satisfy the inequality $\sum_{j=1}^{k} x_{j} \leq \sum_{j=1}^{k} y_{j}$ for all $k\in [n-1]$.  

\begin{figure}[ht!]
\centering
\includegraphics[width=\linewidth]{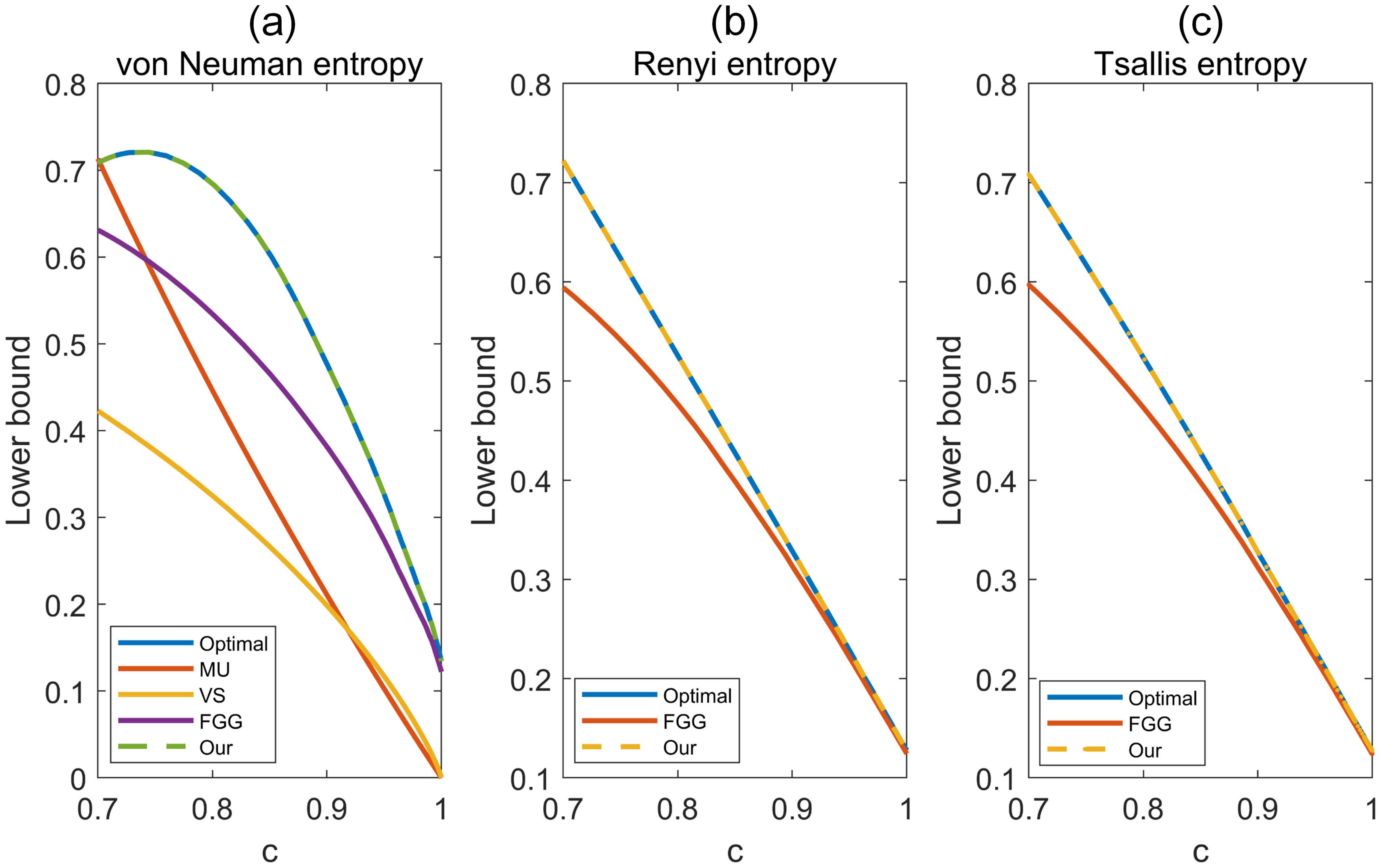}
\caption{Simulations of Entropic Uncertainty Relations on a 2-dimensional Hilbert space. Here, the parameter $c$ denotes the maximal overlap of eigenstates of two observables $A$ and $B$, i.e., $c\equiv \max_{a,b}|\langle \phi_a|\psi_b\rangle|$ with eigenstates $\{\ket{\phi_a}\}$ and $\{\ket{\psi_b}\}$. (a) von Neumann entropy. (b) Renyi entropy ($R_k=\log(\sum_{a}p_a^k)/(1-k)$ with $k=2$). (c) Tsallis entropy ($T_k=(\sum_ap_a^k-1)/(1-k)$ with $k=2$). MU, VS, and FGG denote the lower bound in ref.\cite{Maassen1988}, \cite{Vicente2008}, and \cite{Friedland2013}, respectively.}
    \label{Figure2}
\end{figure}

Since the inequality (\ref{SourceUP}) holds for any uncertainty function $f$, it offers a unified approach to exhibit state-independent uncertainty relations,  as well as coherence and Bell nonlocality (see Sections below). For instance, by using $f(p_a,q_b)=-p_a\log p_a-q_b\log q_b$ with probabilities $p_a,q_b>0$, the inequality (\ref{SourceUP}) leads to the entropic relation as 
\begin{eqnarray}
H(A)+H(B)\geq c_{lb},
\label{Shannon}
\end{eqnarray}
where $c_{lb}=\frac{1}{n}\min_{\rho}\{\sum_{a,b}f(p_a,q_b)\}$ is a constant, and $H(\cdot)$ denotes the Shannon entropy \cite{Cover}. The lower bound $c_{lb}$ typically depends on the maximal overlap of the eigenstates of observable pair \cite{Friedland2013}. Further examples are shown in Appendix using other entropy functions. While analytical bounds may be challenging to obtain, the optimization in Eq. (\ref{SourceUP}) can often be achieved within polynomial complexity using polynomial approximations of nonlinear uncertainty functions \cite{Huber2024}, where all the $d$-dimensional states can be approximated with polynomial number of states for not large enough $d$. Figure~\ref{Figure2} presents three numerical examples that show our bounds outperform previous results \cite{Maassen1988,Vicente2008,Friedland2013}.  

\textbf{Verifying quantum coherence}. As a fundamental principle in quantum physics, Bohr's complementarity principle prohibits a system from exhibiting both wave and particle behaviors simultaneously \cite{Wootters1979,Gisin2002}. The so-called quantum coherence has been investigated with various coherence measures \cite{Baumgratz2014,Streltsov2017,hu2018}. While earlier works focused on complementarity between wave and particle natures \cite{Wootters1979,Gisin2002}, here we connect quantum coherence with our uncertainty relation (\ref{SourceUP}). Specifically, consider an interferometric experiment \cite{Streltsov2017} as a quantum game, shown in Figure~\ref{Figure1}. For two incompatible measurements $A$ and $B$ for Alice and Bob respectively, let $\{p_a, \forall a\}$ and $\{q_b, \forall b\}$ denote their probability distributions. Inspired by the relative entropy coherence measure \cite{Streltsov2017,Lecamwasam2024}, we choose the uncertainty function as $f(p_a,q_b)=-p_a\log p_a+q_b\log q_b$, differing from Schur-concave measures \cite{Masanes2013}. Analogous to the inequality (\ref{SourceUP}), we formulate coherence quantification via a vector-value inequality as
\begin{eqnarray}
\mathbf{0} \prec \btf_{ab}^{_\downarrow} \prec \bR_{\textrm{coh}}(\rho),
\label{coherence}
\end{eqnarray}
where $\mathbf{0}$ denotes the zero vector, $\bR_{coh}(\rho)$ is defined via $R_{k}(\rho)\equiv  \max_{I_{k}}\max_{A}\sum_{(p_a,q_b)\in I_{k}}f(p_a,q_b)$, and $\btf_{ab}^{_\downarrow}$ is defined by $\btf_{ab}^{_{\downarrow}}=[f^{_{\downarrow}}(p_1,q_1), \cdots, f^{_{\downarrow}}(p_n,q_m)]$ corresponding to the arrangement of $\{f(p_a,q_b)\}$ in descending order.  Unlike inequality (\ref{SourceUP}), here state-dependence is crucial because coherence is a property of specific quantum state. The relative entropy coherence \cite{Streltsov2017},  given by $C_r(\rho)=S(\rho_d)-S(\rho)$ (where $\rho_d$ denotes the diagonal matrix of $\rho$), coincides with the relation  (\ref{coherence}), where the lower bound corresponds to incoherent states, while the upper bound follows since von Neumann entropy equals the minimal Shannon entropy over all probability distributions for a given state. Thus, $C_r(\rho)=\max_{A}\{S(\rho_d)-S(\rho)\}=\frac{1}{n}\max_{A,B}\sum_{i,j}f(p_i,q_j)$, where $B$ denotes projection measurement under the computational basis. The inequality (\ref{coherence}) provides a general means to witness quantum coherence using two incompatible measurements. Further extensions for multiple measurements are shown in Appendix.  

\textbf{Verifying Bell nonlocality}. A two-particle system is entangled if it cannot be represented as a mixture of product states \cite{EPR}. So far, there are various methods for detecting entanglement, including entanglement witness, Einstein-Podolsky-Rosen-steering, and Bell inequalities \cite{HHH,Gisin2002,RMPBrunner}, each probing different levels of quantum nonlocal correlations.  Although some results show links between specific entanglement and other quantum properties  \cite{Oppenheim2010,Berta2014,Streltsov2015}, but an explicit connection between measurement incompatibility, coherence, and the full hierarchy of quantum nonlocal correlations  \cite{Jones2007} is still lacking. 

To address this,  we generalize our framework to witness hierarchical quantum correlations by extending  (\ref{SourceUP}) to include both measurement and state variations,  i.e., yielding device-independent uncertainty relations.  Given an uncertainty function in Figure~\ref{Figure1}, defines two device-independent bounds of the $f$-type uncertainty for a given measurement set $\cM_1\times \cM_2$ as 
\begin{eqnarray}
&&R_{k}= \max_{(A, B)\in \cM_1\times \cM_2} R_k(A,B),
\nonumber\\
&&r_{k}=\min_{(A, B)\in \cM_1\times \cM_2} r_k(A,B),
\end{eqnarray}
where $R_{k}(A,B)$ and $r_{k}(A,B)$ are defined in Eq.~(\ref{uuu0}), and $I_{k} \subset [n] \times [m]$. Combining these with the inequality (\ref{SourceUP}) provides uncertainty relations that are independent of both states and measurements: 
\begin{eqnarray}
 \label{dpp}
 \br_{\textrm{sep}}\prec\btf_{ab}\prec \bR_{\textrm{sep}},
\end{eqnarray}
where vectors $\br_{\textrm{sep}}, \btf_{ab} $ and $\bR_{\textrm{sep}}$ are defined similar to Eq.~(\ref{SourceUP}) with $\{r_{k}\}, \{f(p_a,q_b)\}$ and $\{R_{k}\}$, respectively. 

We now apply the inequality (\ref{dpp}) for verifying Bell nonlocality. In a standard Bell experiment \cite{Bell1964}, a source distributes states to space-like separated observers Alice and Bob, who each choose a local measurement ($x$ or $y\in \{0,1\}$) and obtain outcome ($a$ or $b\in \{0,1\}$). Correlations from generalized local hidden variable models \cite{Bell1964} satisfy the Clauser-Horne-Shimony-Holt (CHSH) inequality \cite{CHSH}. Bell nonlocality of entangled states can be verified by violating this inequality \cite{Brunner2014}. We reformulate these scenarios within our game-based method, shown in Figure~\ref{Figure1}. Here, one source sends one particle of a two-particle state. Verifier privately sends each party a measurement setting $x$ (or $y$). Alice and Bob perform measurements according to measurement settings and send outcomes to Verifier. After a large rounds, Verifier can obtain joint conditional probabilities $\{P(a,b|x,y)\}$, and then computes joint uncertainty function $f(P(a,b|x,y))=\sum_{x,y=0,1}(-1)^{xy}P(a,b|x,y)$ (depending on all measurement settings). This differs from standard CHSH game, which assigns payoffs per round \cite{CHSH}. Because the set of separable correlations forms a convex polytope, this setup enables linear optimization for estimating the inequality (\ref{dpp}) via vector majorization. Combining this with the CHSH inequality \cite{CHSH} yields the following inequality,
\begin{eqnarray}
\btf_{ab}^{_{\downarrow}}\prec \mathbf{c},
\label{CHSH}
\end{eqnarray}
where $\btf_{ab}^{_{\downarrow}}$ is defined similar to Eq.~(\ref{SourceUP}) according to the set $\{f(P(a,b|x,y))\}$, $\mathbf{c}=[2,0,0,0]^T$ is the classical bound vector, and T denotes the transpose. The quantum bound is given by $\bc_q=[2\sqrt{2},0,0,0]^T$ \cite{CHSH}. Violation of the relation (\ref{CHSH}), i.e., the vector  $\btf_{ab}^{_{\downarrow}}$ from experimental data is not majorized by $\mathbf{c}$, provides a new method to verify Bell nonlocality. We further extend to entanglement witness and multipartite scenarios in Appendix.

\begin{figure}[!ht]
\centering
\includegraphics[width=\linewidth]{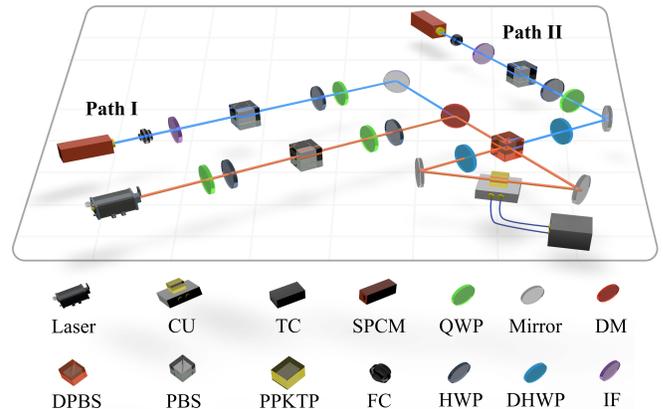}
\caption{Schematic experiment setup. We generated orthogonally polarized photon-pairs at 810 nm wavelength with a PPKTP crystal in a Sagnac interferometer and, by erasing the ``which path'' information of the pump with a dual-wavelength polarization beam splitter (DPBS). Key elements include half-wave plate (HWP), quarter-wave plate (QWP), polarizing beam splitter (PBS), mirror (M), dichromatic mirror (DM), dual-wavelength half-wave plate (DHWP), interference filter (IF), fiber coupler (FC), single-photon counting module (SPCM), control unit (CU), and temperature controller (TC).}
\label{Figure3}
\end{figure}

\textbf{Experimental verifications of entropic uncertainty relations, coherence and Bell nonlocality}. We set up a two-photon experiment to verify the uncertainty relations (\ref{Shannon}),  quantum coherence (\ref{coherence}), and quantum nonlocality by violating the inequality  (\ref{CHSH}). The experimental design and statistical analysis are shown in Appendix.

\begin{figure}[!ht]
\centering
\includegraphics[width=\linewidth]{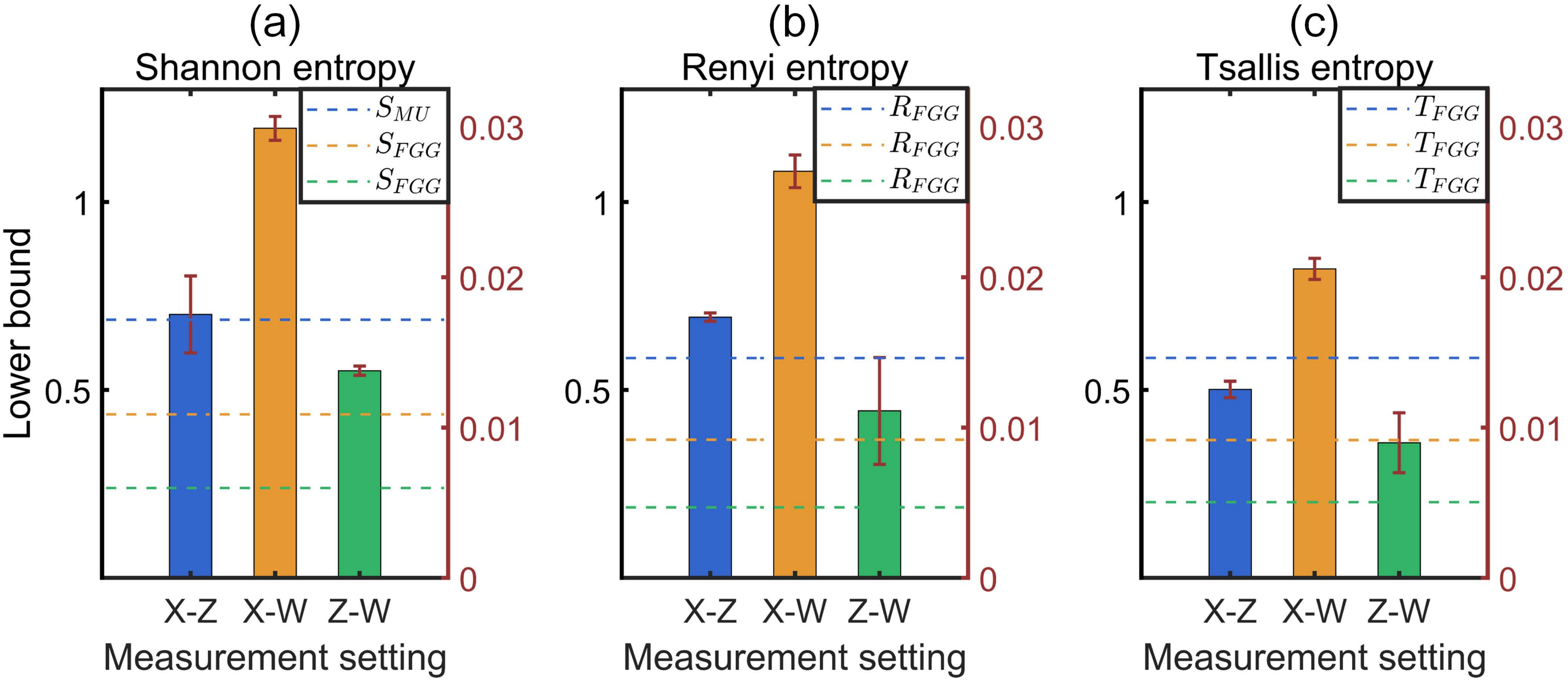}
\caption{Experimental total Shannon entropy (a), Renyi entropy ($q=2$) (b) and Tsallis entropy ($k=2$) (c) of two measurements. Here measurements involve the photon from path I using observables $\{X,Z\}$, and the photon from path II using observables $\{Z,W\}$ with $W=\frac{\sqrt{3}X+Z}{2}$.}
\label{Figure4}
\end{figure}

\begin{figure}[!ht]
\centering
\includegraphics[width=\linewidth]{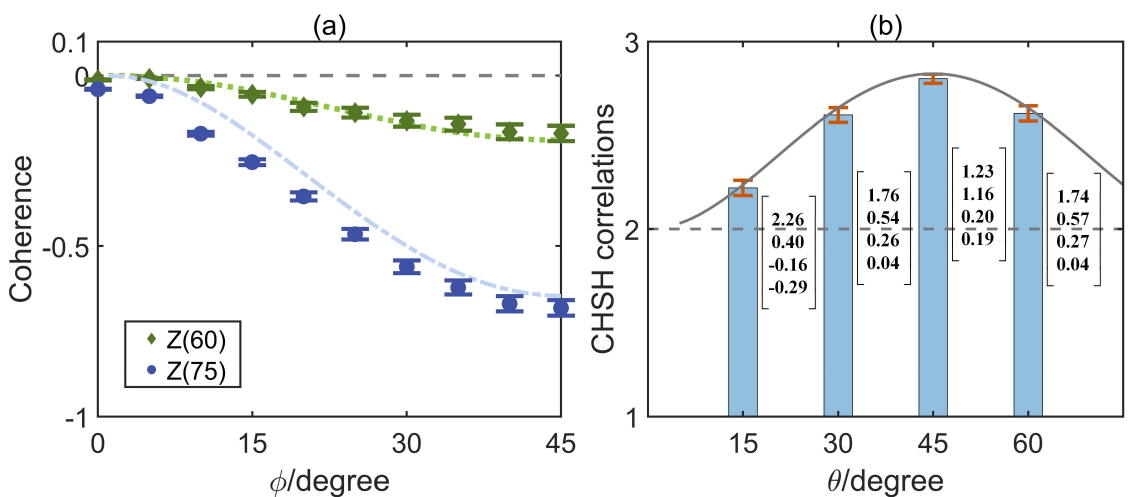}
\caption{Experimental validation of Coherence (a) and 
Bell nonlocality (b). We verified the coherence of the photon in path I for $\theta = 60^\circ$ and $75^\circ$ by performing Pauli $Z$ measurements (denoted as $Z(\theta)$). To demonstrate the nonlocality, we conducted measurements with observables $A_x \in \{Z, X\}$ on the photon in path I and $B_y \in \{\cos\varphi Z +(-1)^y \sin\varphi X\}$ on the photon in path II, where $\varphi$ was chosen to maximize the CHSH correlator.}
\label{Figure5}
\end{figure}

We experimentally validated the entropic uncertainty inequality (\ref{Shannon}) as shown in Figure~\ref{Figure3}. For this purpose, we prepared a set of two-photon entangled states of the form $|\Phi(\theta)\rangle=\sin\theta|HH\rangle+\cos\theta|VV\rangle$, with $\theta\in \Theta:=\{0^\circ$, $10^\circ$, $15^\circ$, $30^\circ$, $45^\circ$, $60^\circ$, $75^\circ$, $90^\circ\}$,  by adjusting the axes of HWP and QWP before the Sagnac interferometer. For each state, we implemented two Pauli measurements along with one superposition: $X$ and $Z$ ($X$-$Z$), $X$ and $W$ ($X$-$W$), and $Z$ and $W$ ($Z$-$W$) with $W:=\frac{\sqrt{3}X + Z}{2}$. The total Shannon entropy $H(\mathbf{a}) + H(\mathbf{b})$ is evaluated based on the majorization of the encoding vector, as described by Eq.~(\ref{SourceUP}). We computed the total Renyi entropy (with $k=2$) and Tsallis entropy (with $k=2$) similarly, as shown in Figure~\ref{Figure4}. 

The results indicate that our measured bounds of Shannon entropy (a) for the measurement settings $X$-$W$ and $Z$-$W$ surpass previous results \cite{Friedland2013}, while for the $X$-$Z$ setting our findings are consistent with existing bounds. For the Renyi entropy (b), our experimental bounds are greater than the previous results \cite{Friedland2013} for all three measurement settings. In the case of Tsallis entropy (c), the experimental bounds are larger than those in ref.\cite{Friedland2013} for the $X$-$W$ and $Z$-$W$ settings, but slightly lower for the $X$-$Z$ setting, likely due to experimental errors. Across all measurements, the experimental data align closely with the ideal values, with errors not exceeding 0.0157. Error bars were estimated using Poissonian counting statistics. This experiment differs from previous demonstrations of variance-based uncertainty relations \cite{Kaneda2014,Zhou2016} and earlier entropic uncertainty relation experiments \cite{nairz2002,pre2011,wang2019}. 

To verify quantum coherence, we prepared the two entangled two-photon states $|\Phi(60^\circ)\rangle$ and $|\Phi(75^\circ)\rangle$, and experimentally confirmed the inequality (\ref{coherence}) using photons from path I. The von Neumann entropy $H(\bb)$ of the diagonal matrix of the density matrix is evaluated via the Shannon entropy of the probability distribution from Pauli $Z$ measurement. For unknown states, the entropy is evaluated using the Shannon entropy associated with measurement in the general orthogonal basis $\{\cos\phi\ket{H}+\sin\phi \ket{V},-\sin\phi\ket{H}+\cos\phi \ket{V}\}$. The maximum difference of two Shannon entropies, defined as $D_H\equiv H(\bb)-H_{\phi}(\ba)$, is found by majorizing the encoding vector $\btf_{ab}$ in accordance with Eq.~(\ref{SourceUP}). For the ideal state from path I, which is a mixture of classical states, i.e., $\rho=\sin^2\theta|H\rangle\langle H|+\cos^2\theta|V\rangle\langle V|$, the entropic coherence should be zero, $S(\rho_d)-S(\rho)=0$. The experimentally determined coherence values are all close to the theoretical prediction, with absolute errors no exceeding 0.0401, see Figure~\ref{Figure5}(a). Error bars are estimated from Poissonian statistics. This experiment validates the coherence-uncertainty relations, and offers a new approach to characterizing quantum non-classicality, distinct from previous studies \cite{silva2016,zheng2018,gao2018,ringbauer2018,karnieli2021,Yang2023}. 

We finally validated Bell nonlocality by experimentally violating the inequality (\ref{CHSH}). This involved preparing four entangled two-photon states $\{\ket{\Phi(15^\circ)}$, $\ket{\Phi(30^\circ)}$, $\ket{\Phi(45^\circ)}$, $\ket{\Phi(60^\circ)}\}$. For each state, we performed two local Pauli measurements $A_x\in \{Z,X\}$ on photons from path I, and local Pauli measurements in the $xz$-plane of the Bloch sphere, $B_y\in \{\cos\varphi{}Z+(-1)^y \sin\varphi{}X\}$, on photon from path II. The maximum bound is determined via majorization of the vector according to Eq.~(\ref{SourceUP}) (Figure \ref{Figure5}(b)). For classical hidden variable models, the upper bound under majorization is $[2,0,0,0]^T$, while for the ideal state $|\Phi(\theta)\rangle$ the maximum violation of the CHSH correlator \cite{CHSH} is given by $\gamma(\theta)\equiv \scriptstyle{2\sqrt{1+\sin^2(2\theta)}}$. This implies the maximum vector $[\gamma(\theta),0,0,0]^T$. In our experiments, four evaluated vectors $\btf_{ab}^{_\downarrow}$ violated the inequality (\ref{CHSH}); specifically, $\btf_{ab}^{_\downarrow}$ is not majorized by $[2,0,0,0]^T$, but is majorized by $[2\sqrt{2},0,0,0]^T$. Error bars are estimated using Poissonian statistics \cite{ap2006,xd2017}, and the $p$-value in the experiment is smaller than $10^{-12}$, ensuring the results of experimental verification. The present experiment provides a new operational explanation of Bell experiments different from previous results \cite{giustina2015,ringbauer2016,moreau2019,Guo2019,ss2023,Chen2024}. We note our experiment does not close either the locality loophole or the measurement loophole, as in the case in most Bell-type experiments \cite{Aerts1999}.

\textbf{Discussions}. The present method provides a unified framework for observing fundamental quantum characteristics, including uncertainty principles, coherence, and Bell nonlocality. In this method, the entropic uncertainty relation, as well as quantum coherence are all interpreted as quantum features that can be witnessed using two incompatible measurements. Demonstrating nonlocality requires a set of incompatible measurements. Moreover, our approach can be extended to tackle the ground-state energy problem. For an $n$-qubit Hamiltonian composed of Pauli matrices, spectral decomposition of all Pauli operators enables the Hamiltonian to be written in terms of semi-positive-definite operators.  Inspired by the domain method \cite{Xu2024}, the upper and lower bounds for ground-state energies correspond to inner and outer approximations of the set of achievable probabilities via local measurements on general states. This leads to a novel uncertainty relation for estimating ground-state energies using polynomial uncertainty functions. Further exploration may yield intriguing insights into quantum many-body systems.

\textbf{Acknowledgements}. This work was supported by the National Natural Science Foundation of China (No. 62172341, No. 12405024, No. 12204386, No. 12371132, and No. 12171044), Sichuan Natural Science Foundation (No. 2024NSFSC1365, No. 2024NSFSC1375),  Fundamental Research Funds for the Central Universities, Sichuan Provincial Key R\&D Program (No. 2024YFHZ0371), Interdisciplinary Research of Southwest Jiaotong University China (No. 2682022KJ004), and the Academician Innovation Platform of Hainan Province.

\appendix

\section{Proof of Inequality (3)}

Under the assumption of independent and identically distributed states being distributed by source, all the uncertainty functions $f(p_a,q_b)$ will be evaluated according to the same state in the given set $\cD$. This implies for any index subset $I_{k} \subset [n] \times [m]$ with given integer $k\in [nm]$ that 
\begin{eqnarray}
  r_{k}(A,B) \leq \sum_{I_k}\sum_{(a,b)\in I_k}f(p_a,q_b),
  \label{Eq:S-1}
\end{eqnarray}
where the quantity $r_{k}(A,B)$ is defined Eq.~(1). Define the vector $\br_{ms}\equiv [r_{1}(A,B), r_{2}(A,B)-r_{1}(A,B), \ldots, r_{mn}(A,B)-r_{mn-1}(A,B)]$. Denote $\btf_{ab}^{_{\uparrow}}=[f^{_{\uparrow}}(p_1,q_1), \cdots, f^{_{\uparrow}}(p_n,q_m)]$ as the arranged vector of $\{f(p_1,q_1), \cdots$, $f(p_n,q_m)\}$ in the increasing order, i.e., $f^{_{\uparrow}}(p_i,q_j)\geq f^{_{\uparrow}}(p_i,q_s)$ for $j\geq s$. Similarly, let $\btf_{ab}^{_{\downarrow}}\equiv [f^{_{\downarrow}}(p_1,q_1), \cdots, f^{_{\downarrow}}(p_n,q_m)]$ be the arranged vector of $\{f(p_1,q_1), \cdots$, $f(p_n,q_m)\}$ in the decreasing order. We obtain from the inequality (\ref{Eq:S-1}) that 
\begin{eqnarray}
\sum_{a=1}^{n_1}\sum_{b=1}^{m_1}f^{_{\uparrow}}(p_a,q_b)&\geq& r_1(A,B)+\sum_{s=2}^\ell (r_{s}(A,B)-r_{s-1}(A,B))
\nonumber
\\
&=&r_{\ell}(A,B),
  \label{Eq:S-2}
  \\
\sum_{a=1}^{n_1}\sum_{b=1}^{m_1}f^{_{\uparrow}}(p_a,q_b)&\leq&\sum_{a=1}^{n_1}\sum_{b=1}^{m_1}f(p_a,q_b)\leq \sum_{a=1}^{n_1}\sum_{b=1}^{m_1}f^{_{\downarrow}}(p_a,q_b),
\nonumber\\
  \label{Eq:S-2a}
\end{eqnarray}
where $\ell=(n_1-1)m+m_1$, $n_1\leq n$ and $m_1\leq m$. From the definition of vector partial order, we obtain the following inequality 
\begin{eqnarray}
\br_{\small \textrm{ms}}\prec\btf_{ab}^{_{\uparrow}} \prec \btf_{ab} \prec \btf_{ab}^{_{\downarrow}}.
\label{Eq:S-5}
\end{eqnarray}
This has proved the inequality (3).

\section{Entropic uncertainty relations}

In this section, we extend entropic uncertainty relations by integrating different entropy measures, including Renyi entropy and Tsallis entropy. Specially, Renyi entropy \cite{Friedland2013} is defined as $R_k(\ba)=\log(\sum_{a}p_a^k)/(1-k)$ ($k>0$ and $k\neq 1$) for a probability distribution $\{p_a\}$. To consider the experiment in Figure 1, define an uncertainty function as $f_R(p_a,q_b)=p_a^kq_b^k$ for each pair of outcomes $a$ and $b$ with the respective probability $p_a$ and $q_b$. The $f$-type uncertainty is bounded from the below by $\min_{\rho\in \cD}f_R(p_a,q_b)$. Define a quantity $r_{k}(A,B)\equiv \min_{I_{k}\subset [n] \times [m]}\min_{\rho\in \cD}\sum_{(a,b)\in I_{k}}f_R(p_a,q_b)$. This leads to a vector-value inequality as
\begin{eqnarray}
\br^{_{\textrm{SI}}}_{_R}\prec \btf^{_{\uparrow}}_{R},
\label{S2:2}
\end{eqnarray}
where the vectors $\br^{_{\textrm{SI}}}_{_R}$ and $\btf^{_\uparrow}_{ab}$ are respectively defined by $\br^{_{\textrm{SI}}}_{_R}=[r_{1}(A,B), r_{2}(A,B)-r_{1}(A,B), \ldots, r_{mn}(A,B)-r_{mn-1}(A,B)]$ and $\btf^{_{\uparrow}}_{R}=[f_R^{_{\uparrow}}(p_1,q_1), \cdots, f_R^{_{\uparrow}}(p_n,q_m)]$ under the increasing order of $\{f_R(p_1,q_1), \cdots, f_R(p_n,q_m)\}$.

As for Tsallis entropy \cite{Tsallis}, it is defined as $T_k(\ba)=(\sum_ap_a^k-1)/(1-k)$ for a probability distribution $\{p_a\}$ with $k>0$ and $k\neq1$. Let the uncertainty function be $f_{T}(p_a,q_b)=p_a^k+q_b^k$ for each pair of outcomes $a$ and $b$ with the respective probability $p_a$ and $q_b$. The $f$-type uncertainty in Figure 1 is bounded from the below by $\min_{\rho\in \cD}f_T(p_a,q_b)$. Using the lower bound $\hat{r}_{k}(A,B)\equiv \min_{I_{k}\subset [n] \times [m]}\min_{\rho\in \cD}\sum_{(a,b)\in I_{k}}f_T(p_a,q_b)$ implies a vector-value inequality as
\begin{eqnarray}
\br^{_{\textrm{SI}}}_{_T}\prec \btf^{_{\uparrow}}_{T},
\label{S2:4}
\end{eqnarray}
where the vectors $\br^{_{\textrm{SI}}}_{_T}$ and $\btf^{_{\uparrow}}_{T}$ are respectively defined by $\br^{_{\textrm{SI}}}_{_T}=[\hat{r}_{1}(A,B), \hat{r}_{2}(A,B)-\hat{r}_{1}(A,B), \ldots, \hat{r}_{mn}(A,B)-\hat{r}_{mn-1}(A,B)]$ and $\btf^{_{\uparrow}}_{T}=[f_T^{_{\uparrow}}(p_1,q_1), \cdots, f_T^{_{\uparrow}}(p_n,q_m)]$ in the increasing order of $\{f_T(p_1,q_1),\cdots, f_T(p_n,q_m)\}$. This provides state-independent uncertainty relations in terms of different entropic functions. 

\section{Universal uncertainty relations with multiple observables}

In this section, we extend to unified uncertainty relations with multiple observables. For the game (Figure 1), consider a given state $\rho$ on a Hilbert space $\cH \cong \mathbb{C}^{d}$. The positive operator-valued measures (POVM) are denoted as $A_{i}= \{M_{a_i}\}$, where $i\in [m]$. For a specific measurement outcome $a_i$, its probability is given by $p(a_i)= \tr(M_{a_i}\rho)$, $a_i\in [n_i]$ and $i\in [n]$. The outcomes $a_i$ are collected into $m$ vectors $\ba_1, \cdots, \ba_m$.

Suppose the uncertainty function $f$ depends on a collection of outcomes $\ba=(a_1,\cdots, a_m)$ with corresponding probabilities $p(a_1), \cdots, p(a_m)$, respectively. Define a quantity as
\begin{eqnarray}
&&r_{k}(p_{a_1},\cdots, p_{a_m})\equiv\min_{I_{k}}\min_{\rho\in\cD}\sum_{{\bf a}\in I_{k}}f(p_{a_1},\cdots, p_{a_m})
\end{eqnarray}
where $I_{k} \subset [n_1] \times \cdots \times [n_m]$ is a subset of $k$ distinct pair of indices, and $k \in [n_1\cdots{}n_m]$. Similar to the proof in Eq.~(\ref{Eq:S-1}), it follows a set of inequalities: $r_{k} (p_{a_1},\cdots, p_{a_m})\leqslant  \sum_{\ba\in I_{k}}f(p_{a_1},\cdots, p_{a_m})$ for $ k\in[N]$ with  $N\equiv n_1\cdots n_m$. This is equivalent to the following vector-valued inequality
\begin{eqnarray}
 \label{dp}
\br^{_{\textrm{SI}}} \prec \btf_{\ba}^{{_\uparrow}} \prec  \btf_{\ba}\prec \btf_{\ba}^{{_\downarrow}},
\label{S3:3}
\end{eqnarray}
where the vector $\br^{_{\rm SI}}$ is defined similar to the inequality (\ref{Eq:S-5}) according to $\{r_i(p_{a_1},\cdots, p_{a_m})\}$. The vector $\btf_{\ba}^{_{\uparrow}}$ denotes the arranged vector of $\btf_{\ba}=\{f(p_{a_1=1},\cdots, p_{a_m=1}), \cdots, f(p_{a_1=m},\cdots, p_{a_m=m})\}$ in the increasing order while the vector $\btf_{\ba}^{_{\downarrow}}$ denotes the arranged vector of $\btf_{\ba}$ in the decreasing order. The inequality (\ref{S3:3}) provides a state-independent uncertainty relation for multiple observables. 

\textit{Example S1}. Consider entropic uncertainty relations with observables $A_1, \ldots, A_m$ as an example. As for the uncertainty function $f_S(p_{a_1},\cdots, p_{a_m})=-\sum_{i=1}^mp_{a_i}\log p_{a_i}$, the inequality (\ref{S3:3}) implies a vector-value uncertainty relation as 
\begin{eqnarray}
    \br_{S}^{_{\textrm{SI}}} \prec\btf_{S}^{\uparrow}.
\end{eqnarray}
This corresponds to the standard sum-based form $\sum_{i=1}^mH(A_i)\geq c'$ in terms of Shannon entropy function and a constant $c'$. Based on the result in ref.\cite{Vicente2008}, we derive the following analytical bound as 
\begin{eqnarray}
\sum_{i=1}^mH(A_i) \geq  \sum_{t=1}^3\sum_{(i,j)\in S_t}h_{ij}^{(t)},
\end{eqnarray}
where $c_{ij}$ denotes the maximal overlaps of two observables $A_i$ and $A_j$, $S_1$, $S_2$ and $S_3$ denote all the measurements such that their maximal overlaps of each pair belongs to $(0,1/\sqrt{2}]$, $[1/\sqrt{2}, 0.834]$ and [0.834, 1], respectively, $h_{ij}^{(1)}=-2\log c_{ij}$ if $0<c_{ij} \leq 1/\sqrt{2}$, $h_{ij}^{(2)}$ if $1/\sqrt{2}\leq c_{ij} \leq  0.834$, and $h_{ij}^{(3)}=-(1+c_{ij}) \log \frac{1+c_{ij}}{2}-(1-c_{ij}) \log \frac{1-c_{ij}}{2}$. 

If using uncertainty function $f_T(p_{a_1},\cdots, p_{a_m})=\sum_{i=1}^mp_{a_i}^k$, we obtain another uncertainty relation as $\br_{T}^{_{\textrm{SI}}} \prec \btf^{_{\uparrow}}_{T}$, which corresponds to the standard sum-based form $\sum_{i=1}^mT_k(A_i)$ in terms of Tsallis entropy. Finally, utilizing function $f_R(p_{a_1},\cdots, p_{a_m})=\prod_{i=1}^mp_{a_i}^k$ implies the following uncertainty relation as 
\begin{eqnarray}
\br_{R}^{_{\textrm{SI}}} \prec  
\btf^{_{\uparrow}}_{R},
\end{eqnarray}
which corresponds to the sum-based uncertainty form in terms of Renyi entropy.

\section{Verifying quantum nonlocality}

In this section, we propose a method for verifying quantum nonlocality by extending the present uncertainty relations.

\subsection{Bell nonlocality}

Suppose a bipartite Bell test involves a source that distributes states to space-like separated observers $\sA$ and $\sB$. The measurement outcome $a$ for $\sA$ depends on local shares and the type of measurement denoted by $x$, while the measurement outcome $b$ for $\sB$ depends on local shares and the type of measurement denoted by $y$. The joint distribution of all outcomes, conditional on measurement settings, is considered nonlocal \cite{Bell1964} if it cannot be decomposed into 
\begin{eqnarray}
P(a,b|x,y)=\int_{\Omega}p(a|x,\lambda)p(b|y,\lambda)d\mu(\lambda),
\end{eqnarray}
where $(\Omega,\mu, \lambda)$ denotes the measure space of hidden variable $\lambda$. In general, all Bell local correlations with given measurement setups constitute polytopes. The Bell correlation polytopes are defined by a set of linear inequalities known as Bell inequalities. The facet inequalities impose constraints on the joint probabilities of measurement outcomes, which are useful for determining whether the observed correlations can be explained by classical or quantum mechanics. If a set of probabilities violates any of the facet inequalities, it indicates the presence of non-classical correlations.

The simplest case, known as the CHSH polytope \cite{CHSH}, is a four-dimensional convex polytope. In general, suppose a bipartite Bell polytope consists of correlations obtained under $d$ measurement settings and $u$ outcomes per party. All classical correlations satisfy the following facet inequality as
\begin{eqnarray}
\sum_{x,y=1}^d\alpha_{xy}E(x,y)\leq c,
\label{S4:2}
\end{eqnarray}
where the correlators are defined by  $E(x,y)=\sum_{a,b=0,1}(-1)^{ab}P(a,b|x,y)$. We can rewrite this Bell inequality into an uncertainty relation. One way is to make use of a nonlocal game similar to it shown in Figure 1. Specially, in each round, one source generates a joint state $\rho_{AB}$ and sends one particle to each quantum observer, Alice and Bob. Each party performs local measurements $\{M_{a|x}\}$ or $\{N_{b|y}\}$ to obtain a single outcome $a$ or $b$ respectively. The uncertainty function used by Verifier depends on joint distributions $\{P(a,b|x,y)\}$ obtained by two quantum observers. 

Define an uncertainty function as $f[P(a,b|x,y)]=\delta_{ab}\sum_{x,y}\alpha_{xy}P(a,b|x,y)$ for two outcomes of all measurement settings. This is different from the CHSH game \cite{CHSH} for each pair of outcomes under given measurement setting. Instead, the function $f(P(a,b|x,y))$ can be regarded as a total uncertainty of all the outcome pairs.  We obtain the following uncertainty relation as 
\begin{eqnarray}
  \btf_{ab} \prec\mathbf{c},
  \label{S4:3}
\end{eqnarray}
where the vector $\btf_{ab}$ is defined according to the arrangement of $\{f[P(a,b|x,y)], \forall a,b\}$ in the decreasing order, and $\mathbf{c}=[c_{11},c_{12},\cdots, c_{dd}]^T$ with $\sum_{a\leq i,b\leq j}f[P(a,b|x,y)]\leq c_{ij}$ and $\sum_{i,j}c_{ij}=c$.

Now, consider a scenario where the source generates an entangled state, denoted as $\rho_{AB}$, on Hilbert space $\cH_A\otimes \cH_B$. Local measurements $\{M_{a|x}\}$ and $\{N_{b|y}\}$ are performed on the particles A and B, respectively. For certain bipartite entanglement, local quantum measurements lead to quantum correlations $\{P(a,b|x,y)\}$ in a standard Bell experiment \cite{Bell1964}, where $P(a,b|x,y)={\rm Tr}(\rho M_{a|x}\otimes N_{b|y} )$. The violation of the inequality (\ref{S4:3}) is equivalent to the existence of a vector $\mathbf{c}_q$ in the nonlocal experiment (Figure 1), violating the inequality (\ref{S4:3}). Here, the violation means that the vector $\mathbf{c}_q$ does not satisfy the inequality (\ref{S4:3}). 

If a source in Bell experiment generates a post-quantum box \cite{PRbox}, a general probability distribution $\{P_{\textrm{ns}}(a,b|x,y)\}$ under the unique constraint of the non-signaling principle violates Bell inequality (\ref{S4:3}). It implies another vector $\bc_{\textrm{ns}}$ in the nonlocal game (Figure 1) violates the inequality (\ref{S4:3}), with the sum of all components potentially surpassing $c_q$ and $c$. This leads to hierarchical uncertainty relations for characterizing quantum nonlocality as 
\begin{eqnarray}
\btf_{ab}\prec\bc \prec \bc_{q} 
\prec \bc_{\textrm{ns}},
\label{S4:5}
\end{eqnarray}
where the vector $\bc_{\textrm{ns}}$ is defined according to the bound by using the non-signaling sources. 

\textit{Example S2}. Consider the nonlocal game depicted in Figure 1. For each pair of two outcomes $(a,b)$ with probabilities $p_a$ and $q_b$, define an uncertainty function $f[P(a,b|x,y)]=\sum_{x,y}(-1)^{xy}p(a|x)q(b|y)(a-\mean{X})(b-\mean{Y})$. Inspired by the covariance-type CHSH inequality \cite{Pozsgay2017}: $\textrm{Cov}(A_0,B_0)+\textrm{Cov}(A_0,B_1)+\textrm{Cov}(A_1,B_0)-\textrm{Cov}(A_1,B_1)\leq \frac{16}{7}$, we derive an uncertainty relation to characterize a set of measurements performed on any separable states as $\btf_{ab}^{_\downarrow} \prec \mathbf{c}$, where the vector $\mathbf{c}$ is defined by $\mathbf{c}=(4/7,4/7,4/7,4/7)^T$. Here, $Cov(A,B)$ denotes the covariance matrix of two observables. This provides a device-independent uncertainty relation to witness the macroscopic nonlocality \cite{Watts2021}. 

\subsection{Entanglement Witness}

Different from the device-independent witness of entanglement \cite{Bell1964}, any entangled state can be verified by using the entanglement witness operator \cite{HHH}, where all separable states form a convex set. Specially, for a given entangled state $\rho$ on Hilbert space $\mathcal{H}_{AB}$, there exists an entanglement witness operator $\mathcal{E}$ satisfying that $\tr(\rho \mathcal{E})>0$ and $\tr(\rho_{s} \mathcal{E})\leq 0$ for any separable state $\rho_s$ on $\mathcal{H}_{AB}$. Suppose that the observable-based decomposition of $\mathcal{E}$ is given by $\mathcal{E}=\sum_{x,y=1}^{\ell}\alpha_{xy}A_x\otimes B_y$ with local quantum observables $A_x$ and $B_y$ and some constants $\alpha_{xy}$. Let $\{M_{a|x}\}$ and $\{N_{b|y}\}$ be eigenstates associated with the eigenvalues $\mu_{a}$ and $\gamma_b$ of quantum obversables $A_x$ and $B_y$ respectively, where $\{M_{a|x}\}$ and $\{N_{b|y}\}$ satisfy  $\sum_aM_{a|x}=\sum_bN_{b|y}=\mathbbm{1}$ for any $x,y$. We obtain another decomposition as 
\begin{eqnarray}
\mathcal{E}=\sum_{x,y=1}^{\ell}\sum_{a,b=1}^d\alpha_{xy}\mu_{a}\gamma_{b}M_{a|x}\otimes N_{b|y}.
\end{eqnarray}
Denote the joint probability $P(a,b|x,y)=\tr(\rho M_{a|x}\otimes N_{b|y})$. The entanglement witness can then be regarded as a measurement-dependent uncertainty relation similar to Eq.~(\ref{S4:2}) with respect to a given set of measurements $\{M_{a|x}\otimes N_{b|y}, \forall a,b\}$ with $x,y\in[d]$ as 
\begin{eqnarray}
\btf_{ab}\prec \mathbf{c}_s,
\label{S4:7}
\end{eqnarray}
where the uncertainty function is defined by $f[P(a,b|x,y)]=\mu_{a}\gamma_b\sum_{x,y}\alpha_{xy}P(a,b|x,y)$, and $\mathbf{c}_s$ denotes the upper bound over all separable states generated by Verifier in Figure 1 and satisfies the summation of all components is zero. Violating the inequality means there is a vector that does not satisfy the inequality (\ref{S4:7}). Specially, for an entangled state $\rho$, denote $\bc_q (\rho)$ as the vector obtained by measuring the entangled state $\rho$. The vector $\bc_q (\rho)$ satisfies the summation of all components is $c_q(\rho)=\tr(\rho E)>0$. This can be extended by using the EPR-steering \cite{EPR,He2013}, and quantum contextuality \cite{Pavicic2023}.

\subsection{Multipartite scenarios}

In an $n$-partite Bell test, a source distributes states to space-like separated observers $\sA_1, \cdots, \sA_n$ \cite{Bell1964}. The measurement outcome $a_i\in \{1,\cdots, d\}$ of observer $\sA_i$ depends on local shares and the type of measurement denoted by $x_i$. The joint distribution of all outcomes, conditional on measurement settings, is considered genuinely multipartite nonlocal if it cannot be decomposed as follows \cite{Svetlichny1987}: 
\begin{eqnarray}
P_{\textrm{bs}}(\ba|\bx)
=\sum_{I,\oI}p_{I,\oI}P(\ba_I|\bx_I) P(\ba_{\oI}|\bx_{\oI}), 
\end{eqnarray}
where $\bu=(u_1,\cdots, u_n)$ with $u\in \{a,x\}$, $\bu_{J}:=(u_i,i\in J)$, $a_i$ and $x_i$ denote inputs and outcomes per party, $I$ and $\oI$ define a bipartition of $\{1, \cdots, n\}$, and $\{p_{I,\oI}\}$ is a probability distribution over all bipartitions. The joint distribution $P(\ba_I|\bx_I)$ is contingent on the measurement settings $\bx_I$. Similar notation is employed for $P(\ba_{\oI}|\bx_{\oI})$.

The multipartite correlation is termed as \textit{non-signaling biseparable} if both $P(\ba_I|\bx_I)$ and $P(\ba_{\oI}|\bx_{\oI})$ can exhibit any non-signaling correlations \cite{PRbox}. It is considered  \textit{nonlocal biseparable} if they represent quantum correlations. This concept can be extended for biseparable entanglement \cite{HHH} or the biseparable EPR-steering \cite{EPR,He2013}.  In contrast, fully separable correlations in Bell scenarios are classical correlations from hidden variable models \cite{Bell1964}.

Similar to bipartite scenarios, multipartite correlations can be represented by correlation polytopes. The facets of these polytopes may correspond to classical or biseparable correlations. Analyzing these polytopes enables the discrimination between classical and various levels of multipartite quantum behaviors in Bell experiments. In particular, consider a multipartite Bell polytope with $d$ measurement settings and $u$ outcomes per party. All classical or biseparable (local or short) correlations satisfy the following facet inequality:
\begin{eqnarray}
\sum_{\bx}\alpha_{\bx}E(\bx)\leq c.
\label{S4:9}
\end{eqnarray}
Here, the correlators $E(\bx)=\sum_{\ba}\delta_{\ba}
P(\ba|\bx)$ represents the correlations of all the measurement outcomes for a given collection of measurement settings $\bx$, and parameters $\delta_{\ba}$ depend on the measurement outcomes, such as $\delta_{\ba}=(-1)^{\prod_ja_j}$ when $a_1,\cdots, a_n\in\{0,1\}$.

Suppose one source in Figure 1 generates an $n$-particle state $\rho_{A_1\cdots{}A_n}$ on Hilbert space $\cH_{A_1}\otimes \cdots\otimes \cH_{A_n}$. Define $\{M_{a_i|x_i}\}$ as local measurements being preformed on particle $A_i$, $i\in [n]$. Here, it does not require all quantum players to receive the same reduced state. The joint distribution is given by $P(\ba|\bx)=\tr(\rho M_{a_1|x_1}\otimes \cdots \otimes M_{a_n|x_n})$. Define the certainty function as $f[P(\ba|\bx)]=\delta_{\ba}\sum_{\bx}\alpha_{\bx}P(\ba|\bx)$ for a collection of measurement outcomes $\ba$. We can express the Bell inequality (\ref{S4:9}) in terms of a device-independent uncertainty relation with respect to a set of local measurements on separable states:
\begin{eqnarray}
\btf_{\ba} \prec\mathbf{c},
\label{S4:10}
\end{eqnarray}
where the vector $\btf_{\ba} $ is defined according to the quantity $f(\ba)$, and the vector $\bc=[c_{1\cdots{}1},\cdots,c_{d\cdots{}d}]^T$ considering all fully separable states. 

Regarding multipartite quantum states, there exist other vectors that violate the inequality (\ref{S4:10}), such as $\bc_{\textrm{bs}}$, $\bc_{\textrm{gem}}$ and $\bc_{\textrm{ns}}$, representing the upper bound over biseparable states, genuine entanglement, or a post-quantum source, respectively. This implies hierarchic uncertainty relations as 
\begin{eqnarray}
\btf_{\ba}
\prec \bc \prec \bc_{\textrm{bs}} \prec \bc_{\textrm{gem}} \prec \bc_{\textrm{ns}}.
\label{S4:11}
\end{eqnarray}
Similar results can be extended using multipartite entanglement witness \cite{HHH} or multipartite EPR steering \cite{He2013}.

\textit{Example S3}. Consider the Svetlichny inequality \cite{Svetlichny1987} as 
\begin{eqnarray}
S_3=\sum_{i,j,k=0,1}\delta_{ijk}A_iB_jC_k\leq 4,
\end{eqnarray}
where $\delta_{ijk}=-1$ if and only if $i=j=k$, and 1 otherwise, and the quantum bound is $4\sqrt{2}$ and the non-signaling bound is $8$. This can be rewritten into the following vector-valued uncertainty relation as \begin{eqnarray}
\btf_{abc}\prec \bc_{v} \prec \bc_{q}=\bc_{\textrm{ns}},
\end{eqnarray}
where the function $f$ is defined as $f[P(a,b,c|x,y,z)]=\sum_{x,y,z=0,1}\Lambda_{xyz}P(a,b,c|x,y,z)$ for three outcomes $(a,b,c)$, $\bc=[4,0,0,0,0,0,0,0]^T$, $\bc_{q}=[4\sqrt{2},0,0,0,0,0,0,0]^T$ and $\bc_{\textrm{ns}}=[8,0,0,0,0,0,0,0]^T$. 

\section{Experiments}

\subsection{Statistical Analysis}

To experimentally validate the universal uncertainty relations, we generate two-photon entangled states $|\Phi(\theta)\rangle$ by adjusting the parameter $\theta$ through the axis direction of HWP and QWP before the Sagnac interferometer (Figure 2). A collection of two-photon states is created, varying $\theta$ across the set $\Theta:=\{0^\circ, 10^\circ$, $15^\circ, 30^\circ$, $45^\circ, 60^\circ, 75^\circ, 90^\circ\}$. For numerical optimization and density operator reconstruction, we employ different polarization projections following the method \cite{James}. 

Local projection measurement bases are constructed for both parties, denoted as $\{M_{a|x}\equiv |\varphi_x^a\rangle \langle\varphi_x^a|\}$ for one party and $\{N_{b|y}\equiv |\varphi_y^b\rangle \langle\varphi_y^b|\}$ for the other. Measurements $\{M_{a|x}\}$ are performed on the photon in path I, and $\{N_{b|y}\}$ measurements on the photon in path II, where both measurement bases are chosen in accordance with the verified uncertainty relation. For the input state $\ket{\Phi(\theta)}$ with $\theta\in \Theta$, the two-photon joint probability $P_\theta(a,b|x,y)$ of outcomes $a, b\in \{0,1\}$ conditional on measurement settings $x$ and $y$ is evaluated by  
\begin{eqnarray}
P_\theta(a,b|x,y)=\frac{N_{x,y}^{a,b}}{N_{x,y}^{0,0}+N_{x,y}^{0,1}+N_{x,y}^{1,0}+N_{x,y}^{1,1}},
\end{eqnarray}
where $N_{x,y}^{a,b}$ denotes the photon coincidence number. The density matrix of the two-photon state is estimated through state tomography with the post-selection of 100 sets of photon coincidences. The fidelity between two states $\rho_1$ and $\rho_2$ is defined by $F(\rho_1, \rho_2)=(\mathrm{Tr}\sqrt{\sqrt{\rho_1}\rho_2\sqrt{\rho_1}})^2$ \cite{HHH}.

To verify universal entropic uncertainty relations, the uncertainty function $f_\theta(p_a,q_b)=-p_\theta(a|x)\log p_\theta(a|x)-p_\theta(b|y)\log p_\theta(b|y)$ for each pair of outcomes $a$ and $b$ is computed using the marginal probability distributions $p_\theta(a|x)=\sum_{b}P_\theta(a,b|x,y)$ and $p_\theta(b|y)=\sum_{a}P_\theta(a,b|x,y)$. The total Shannon entropy of measurement statistics is evaluated through the vector partial order of the certainty function according to Eq.~(4). Similarly, the total Renyi entropy with $q=2$ is computed using the certainty function $f_\theta(p_a,q_b)=1-p_\theta(a|x)^2-p_\theta(b|y)^2$. 

To verify the coherence uncertainty relations, we evaluate the certainty function $f_{\theta;\phi}(p_a,q_b)=-p_\theta(b)\log p_\theta(b)+ p_\theta(a|x_\phi)\log p_\theta(a|x_\phi)$ for each pair of outcomes and all measurement settings using the marginal probability distributions $p_\theta(a|x_\phi)$ and $p_\theta(b)$.

To verify Bell nonlocality, we evaluate the certainty function $f[P(\ba|\bx)]=\sum_{x,y=0,1}(-1)^{xy}P_\theta(a,b|x,y)$ using the joint probability distributions $P_\theta(a,b|x,y)$. All the maximal quantities are evaluated through the partial order of the encoding vector. In statistical analysis, $p$-values are estimated under the Poissonian distribution, which does not rely on assumptions of Gaussian distribution or the independence of each trial result.

\subsection{Experimental Data}

In this section, we present the experimental data used in Figures 4 and 5. This includes quantum interference, quantum state tomography, fidelities, marginal probabilities of the measurement outcomes of each qubit for evaluating the entropic uncertainty relations, the marginal probabilities of the measurement outcome for evaluating the coherence of one qubit, and CHSH correlations of CHSH test. 

\begin{figure}[!ht]
\centering
\includegraphics[width=\linewidth]{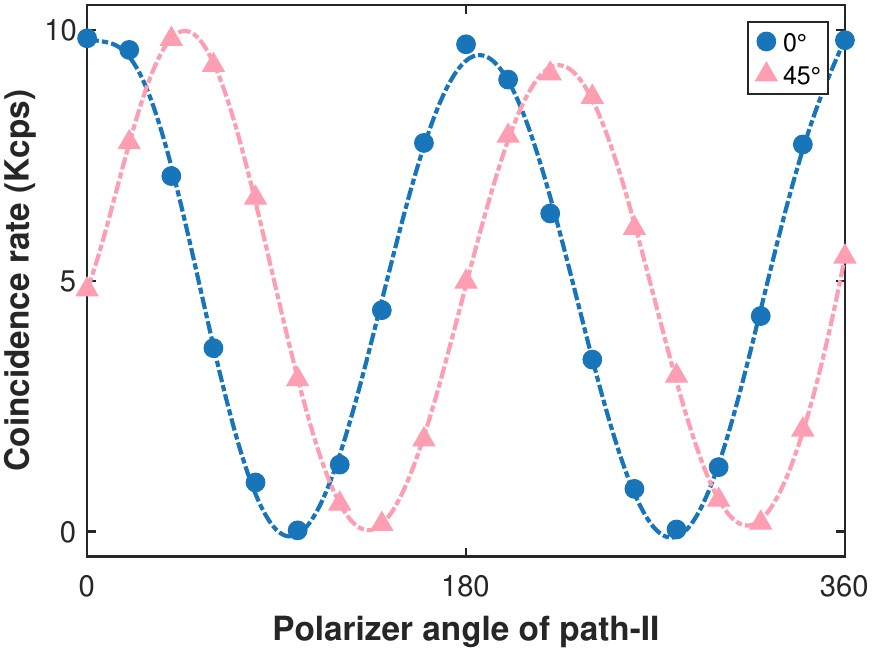}
\caption{Interference. Interference patterns of the two-photon maximally entangled state $\ket{\Phi(45^\circ)}=\frac{1}{\sqrt{2}}(\ket{HH}+\ket{VV})$ from path I with $\phi_{\rm{pol}}=0^\circ$ and $\phi_{\rm{pol}}=45^\circ$ are shown.}
\label{FigureS4}
\end{figure}

The experimental results for the coincidence rate of the two-photon maximally entangled state $\ket{\Phi(45^\circ)}=\frac{1}{\sqrt{2}}(\ket{HH}+\ket{VV})$ are plotted against the polarizer angle of path-II, with the polarizer angle of path-I fixed at 0 and $\pi/4$ degrees. The result aligns with the theoretical quantum predictions, as shown in Figure \ref{FigureS4}.

\begin{figure*}[!ht]
\centering
\includegraphics[width=0.9\linewidth]{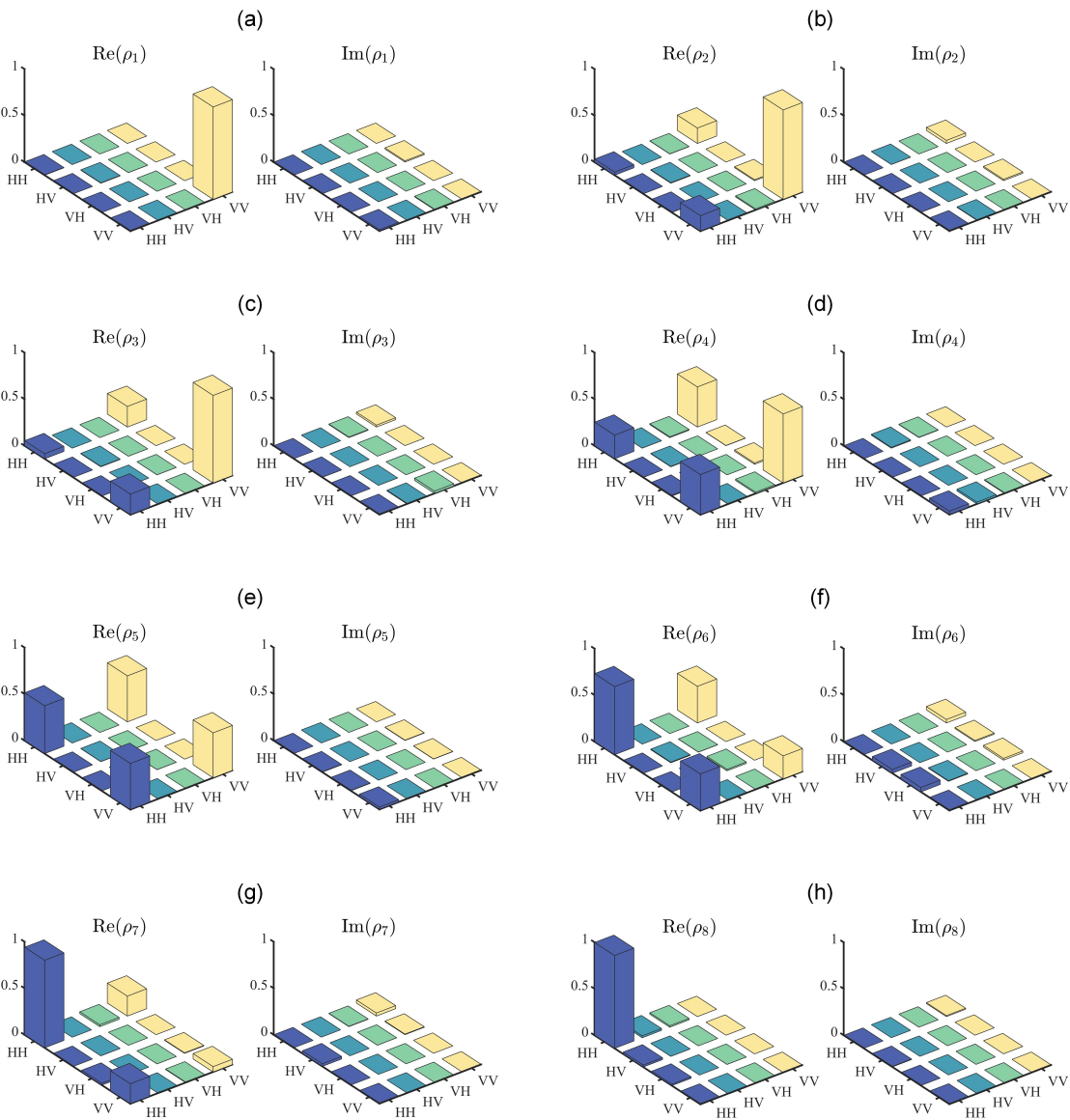}
\caption{The tomographic results for the eight states as $|\Phi(\theta)\rangle=\sin\theta|HH\rangle+\cos\theta|VV\rangle$ with $\theta\in \Theta:=\{0^\circ, 10^\circ$, $15^\circ$, $30^\circ$, $45^\circ$, $60^\circ, 75^\circ,90^\circ\}$. (a) $\rho_1=|\Phi(0^\circ)\rangle \langle\Phi(0^\circ)|$. (b) $\rho_2=|\Phi(10^\circ)\rangle \langle\Phi(10^\circ)|$. (c) $\rho_3=|\Phi(15^\circ)\rangle \langle\Phi(15^\circ)|$. (d) $\rho_4=|\Phi(30^\circ)\rangle \langle\Phi(30^\circ)|$. (e) $\rho_5=|\Phi(45^\circ)\rangle \langle\Phi(45^\circ)|$. (f) $\rho_6=|\Phi(60^\circ)\rangle \langle\Phi(60^\circ)|$. (g) $\rho_7=|\Phi(75^\circ)\rangle \langle\Phi(75^\circ)|$. (h) $\rho_8=|\Phi(90^\circ)\rangle \langle\Phi(90^\circ)|$. The ``Re'' represents the real part of the density matrices and the ``Im'' for the imaginary part.}
\label{FigureS3}
\end{figure*}

We utilize state tomography to reconstruct the density matrix of the two photons across 100 sets of photon coincidences in Figure \ref{FigureS3}(a)-(h). The results demonstrate the fidelity of the prepared two-photon states is larger than 0.98 for all states in Table \ref{Table3 Fidelity}. The two reduced states of the single photon exhibit a similar density matrix, ensuring the validity of the assumption of two states. Specially, the fidelity of the two reduced density matrices, $\rho_A=\tr_B\rho_{AB}$ and $\rho_B=\tr_A\rho_{AB}$, is greater than 0.99 based on experimental evaluations.

We experimentally validated the entropic uncertainty inequality (4), and the marginal probabilities are shown in Table \ref{Tabpro}. The state is prepared as $|\Phi(\theta)\rangle = \sin\theta|HH\rangle + \cos\theta|VV\rangle$, where $\theta$ is chosen from $\{0^\circ, 10^\circ, 15^\circ, 30^\circ, 45^\circ, 60^\circ, 75^\circ, 90^\circ\}$.

To verify the relation (5) for quantum coherence, we utilized the photons from the path I of the two entangled states $|\Phi(60^\circ)\rangle$ and $|\Phi(75^\circ)\rangle$. The marginal probabilities of the measurement outcomes used to evaluate the coherence of one qubit are presented in Table \ref{Tabcoh}.

Table \ref{TabCHSH} shows the experimental coincidence records of CHSH tests for the four entangled two-photon states $\{\ket{\Phi(15^\circ)}$, $\ket{\Phi(30^\circ)}$, $\ket{\Phi(45^\circ)}$, $\ket{\Phi(60^\circ)}\}$ by demonstrating a violation of the inequality (8).

\begin{table*}[hp!]
    \centering
        \caption{Fidelity of two two-photon and single-photon states. Here, the fidelity of two states $\rho_i$ is defined by $F(\rho_1,\rho_2)=(\tr{\sqrt{\sqrt{\rho_1}\rho_2\sqrt{\rho_1}}})^2$.}     
    \newcolumntype{X}{>\raggedleft\arraybackslash}    
    \begin{tabularx}{\textwidth}{X|XXXXXXXXX}
    \hline\hline
\hspace{0.5cm}$\theta$\hspace{0.25cm}  & $0^\circ$ & $10^\circ$ & $15^\circ$ & $30^\circ$ & $45^\circ$ & $60^\circ$ &  $75^\circ$ & $90^\circ$  \\
       \hline
       Two-qubit states&  0.9926  &  0.9926 &   0.9918  &  0.9928    &0.9930 &   0.9928 & 0.9890   & 0.9980  \\
       One-qubit states& 0.9976 & 0.9976 & 0.9992 & 0.9996 & 0.9998 & 0.9986 & 0.9984 & 0.9996  \\
       \hline\hline
        \end{tabularx}        
    \label{Table3 Fidelity}
\end{table*}

\begin{table*}[hp!]
    \centering
          \caption{The marginal probabilities for verifying entropic uncertainty relations. We present marginal probabilities of the measurement outcomes $a$ and $b$. The Shannon entropy $H(\ba)+H(\bb)$ is evaluated by using quantum basis $H/V:=\{\ket{H},\ket{V}\}$, $D/A:=\{\frac{1}{\sqrt{2}}(\ket{H}\pm \ket{V})\}$ and $G/K:=\{\frac{1}{2}(\ket{H}-\sqrt{3}\ket{V}), \frac{1}{2}(\sqrt{3}\ket{H}+\ket{V})\}$. The state is prepared as the $\sin\theta|HH\rangle+\cos\theta|VV\rangle$. }
    {\fontsize{10}{10}\selectfont
\begin{ruledtabular}
\begin{tabular}{c|ccccccccc}
 \multicolumn{3}{c}{} & \multicolumn{4}{c}{$D/A$ and $H/V$} \\
    \hline
        $\theta$ & $0^\circ$ & $10^\circ$ & $15^\circ$ & $30^\circ$ & $45^\circ$ & $60^\circ$  & $75^\circ$ &  $90^\circ$ \\
         \hline
    $p(a=0)$ & 0.0074  & 0.0401  & 0.0579  & 0.2600 
             & 0.5160  & 0.7403  & 0.9392  & 0.9991  \\
    
    $p(a=1)$ & 0.9928  & 0.9599  & 0.9434  & 0.7440 
             & 0.4882  & 0.2597  & 0.0557  & 0.0011    \\
    
    $p(b=0)$ & 0.4914  & 0.5005  & 0.4795  & 0.5189  
             & 0.4522  & 0.4682  & 0.4994  & 0.5233 \\
    
    $p(b=1)$ & 0.5102  & 0.4973  & 0.5169  & 0.4901 
             & 0.5467  & 0.5337  & 0.5008  & 0.4815 \\
\hline
 \multicolumn{3}{c}{} & \multicolumn{4}{c}{$D/A$ and $G/K$} \\
 \hline
    $p(a=0)$ & 0.2018  & 0.2444  & 0.2516  & 0.3883 
             & 0.4689  & 0.5631  & 0.7357  & 0.7623  \\
    
    $p(a=1)$ & 0.7982  & 0.7556  & 0.7484  & 0.6117 
             & 0.5311  & 0.4369  & 0.2643  & 0.2377    \\
    
    $p(b=0)$ & 0.4862  & 0.5015  & 0.4823  & 0.5154  
             & 0.4542  & 0.4674  & 0.4985  & 0.5203 \\
    
    $p(b=1)$ & 0.5138  & 0.4985  & 0.5177  & 0.4846 
             & 0.5458  & 0.5326  & 0.5015  & 0.4797 \\
\hline
 \multicolumn{3}{c}{} & \multicolumn{4}{c}{$H/V$ and $G/K$} \\
\hline
    $p(a=0)$ & 0.0069  & 0.0386  & 0.0586  & 0.2581 
             & 0.5158  & 0.7380  & 0.9422  & 0.0578  \\
    
    $p(a=1)$ & 0.9931  & 0.9614  & 0.9414  & 0.7419 
             & 0.4842  & 0.2620  & 0.0578  & 0.0011    \\
    
    $p(b=0)$ & 0.2397  & 0.2681  & 0.2620  & 0.3904 
             & 0.4678  & 0.5951  & 0.7196  & 0.7670 \\
    
    $p(b=1)$ & 0.7603  & 0.7319  & 0.7380  & 0.6096 
             & 0.5322  & 0.4049  & 0.2804  & 0.2330 \\
    \end{tabular}
    \label{Tabpro}
\end{ruledtabular}
}
\end{table*}

\begin{table*}[ht!]
    \centering
 \caption{The marginal probabilities for verifying coherence. We present the marginal probabilities of the measurement outcome $a$ to evaluate coherence. We evaluate the quantity $H(\bb)-\min_{\phi}H(\ba)$ by using quantum basis  $\{\cos\phi\ket{H}+\sin\phi \ket{V},-\sin\phi\ket{H}+\cos\phi \ket{V}\}$ on the photon in Path I of the joint state $|\Phi(\theta)\rangle$.}
 \setlength{\tabcolsep}{2pt}
   {\fontsize{10}{10}\selectfont
\begin{ruledtabular}
\begin{tabularx}{\textwidth}{c|p{0.8cm} ccccccccc}
 \multicolumn{3}{c}{} & \multicolumn{5}{c}{$\theta=60^{\circ}$} \\
   \hline
        $\phi$ & $0^\circ$ & $5^\circ$& $10^\circ$& $15^\circ$& $20^\circ$& $25^\circ$& $30^\circ$& $35^\circ$& $40^\circ$& $45^\circ$\\    
        \hline
       $p(a=0)$ & 0.7293&    0.7321&    0.7127  &  0.6971 &   0.6632  &  0.6447 &   0.6141 &   0.5981&    0.5429   & 0.5167 \\
        $p(a=1)$ & 0.2707 &   0.2679 &   0.2873 &   0.3028 & 0.3368 &   0.3553 &   0.3859 &   0.4018 & 0.4571  &  0.4833   \\
   \hline
 \multicolumn{3}{c}{} & \multicolumn{5}{c}{$\theta=75^\circ$} \\
   \hline
        $p(a=0)$ & 0.9319  &  0.9264 &   0.8937 &   0.8645 &   0.8233  &  0.7665 &   0.7020 &   0.6434 &   0.5661 &   0.4889 \\
        $p(a=1)$ & 0.0681  &  0.0736 &   0.1063  &  0.1355 &   0.1768  &  0.2335   & 0.2980 &   0.3566 &   0.4339  &  0.5110 \\
    \end{tabularx}
    \end{ruledtabular}
    }
    \label{Tabcoh}
\end{table*}

\begin{table}[ht!]
\centering
\caption{Experimentally coincidence record of CHSH test. Here, all the tests are performed on the joint state $|\Phi(\theta)\rangle$ with $\theta=15^{\circ}$, $30^{\circ}$, $45^{\circ}$ and $60^{\circ}$.}
\begin{ruledtabular}
\begin{tabular}{cccccccc}
 \multicolumn{2}{c}{Input} & \multicolumn{2}{c}{Output} & \multicolumn{4}{c}{Coincidence Rate [cps]} \\
 \colrule
 $x$ & $y$ & $a$ & $b$ & $\theta=15^{\circ}$ & $\theta=30^{\circ}$ & $\theta=45^{\circ}$ & $\theta=60^{\circ}$  \\
 \colrule
 \multirow{4}{*}{0} & \multirow{4}{*}{0} & 0 & 0   & 14560 & 11716  & 26058 & 4046   \\
  & & 0 & 1 & 630   &  1602 & 4259 & 556  \\
  & & 1 & 0 & 139   & 753  & 4936 & 1551  \\
  & & 1 & 1 & 1047  & 4061 & 27230 & 11393  \\
 \colrule
  \multirow{4}{*}{0} & \multirow{4}{*}{1} & 0 & 0   & 14581 & 12066 & 26470 & 3927   \\ 
  & & 0 & 1   & 856 & 1715 & 4342 & 695  \\
  & & 1 & 0   & 49 & 575 & 4699 & 1608  \\
  & & 1 & 1   & 1121 & 4313  & 28175 & 10792  \\
 \colrule
  \multirow{4}{*}{1} & \multirow{4}{*}{0} & 0 & 0   & 8470 & 8948 & 24472 & 5043   \\
  & & 0 & 1   & 95 & 422  & 5044 & 3436  \\
  & & 1 & 0   & 6546 & 3480 & 4905 & 487  \\
  & & 1 & 1   & 1739 & 5402 & 26020 & 8117  \\
  \colrule
  \multirow{4}{*}{1} & \multirow{4}{*}{1} & 0 & 0   & 75 & 392 & 5017 & 3051  \\
  & & 0 & 1   & 8524 & 8883 & 25284 & 5181  \\
  & & 1 & 0   & 1840 & 5769 & 27045 & 8011  \\
  & & 1 & 1   & 6684 & 3979 & 4584 & 389  \\
\end{tabular}
\end{ruledtabular}
\label{TabCHSH}
\end{table}

\end{document}